\documentclass[12pt]{article}
\usepackage{amssymb}
\usepackage{amsmath}
\title{Comment on Mente, O'Donnell, Rangarajan, {\it et al.\/} ``Associations of urinary sodium excretion with cardiovascular events in individuals with and without hypertension: a pooled analysis of data from four studies''}
\author{J. I. Katz \\ Dept. Physics, Washington University St. Louis, Mo. 63130}
\date{\today}
\begin{document}
\maketitle
\begin{abstract}
Mente, O'Donnell, Rangarajan, {\it et al.\/} ignore a possible source of bias
that may invalidate their finding of an anticorrelation between sodium intake
and cardiovascular events.
\end{abstract}

Mente, O'Donnell, Rangarajan, et al. \cite{MOR16} ignore a possible source of bias in their analysis of association of sodium excretion with cardiovascular events.  The population at any level of blood pressure will comprise individuals with varying levels of sodium sensitivity, and those with higher sensitivity will achieve that blood pressure if they ingest less sodium.   As a result, within any blood pressure level, for example, normotension, there will be an anticorrelation between sodium intake and sodium sensitivity.

If sodium sensitivity is causal of cardiovascular events or death this could explain the observational anticorrelation between sodium intake and these events.  Such a causal relation between sodium sensitivity and adverse outcomes is plausible: sensitive individuals are likely to have been hypertensive in the past and to be so in the future.  As a result, the observational association of low sodium excretion with cardiovascular events may not be causal, but an artefact of the anticorrelation between sodium intake and sensitivity within the normotensive group.

\end{document}